\documentclass[conference]{IEEEtran}
\usepackage{amsmath}
\usepackage{amssymb}
\usepackage{amsfonts}
\usepackage{graphicx}
\usepackage{epsfig}
\usepackage{subfigure}
\usepackage{psfrag}
\usepackage{epstopdf}
\usepackage{graphicx,color,psfrag,subeqnarray}
\usepackage{cite}
\usepackage{latexsym}
\usepackage{url}
\usepackage{color}
\usepackage{bigstrut}
\usepackage{multirow}
\usepackage{authblk}
\usepackage{bm,array}
\usepackage{array}
\DeclareGraphicsExtensions{.pdf,.jpeg,.png,.jpg}
\usepackage{lipsum}
\begin{document}
\title{Energy Management for Demand Responsive Users with Shared Energy Storage System}
\author[1]{Katayoun Rahbar}
\author[2]{Mohammad R. Vedady Moghadam}
\author[1,2]{Sanjib Kumar Panda}
\author[1]{Thomas Reindl}
\affil[1]{Solar Energy Research Institute of Singapore (SERIS). E-mail: \{serkr,thomas.reindl\}@nus.edu.sg}
\affil[2]{ECE Department, National University of Singapore (NUS). E-mail: \{elemrvm,sanjib.kumar.panda\}@nus.edu.sg}
\maketitle
\thispagestyle{empty}

\begin{abstract}
This paper investigates the energy management problem for multiple self-interested users, each with renewable energy generation as well as both the fixed and controllable loads,  that all share a common energy storage system (ESS). 
The self-interested users are willing to  sell/buy  energy  to/from  the shared ESS if they can achieve lower energy costs compared to the case of no energy trading  while preserving their privacy e.g. sharing only limited information with a central controller. 
Under this setup,  we propose an iterative algorithm by which the central controller coordinates the charging/discharging values to/from the shared ESS by all users such that their individual energy costs reduce at the same time. 
For performance benchmark, the case of cooperative users that all belong to the same entity is considered, where they share all the required information with the central controller so as to minimize their total energy cost. Finally, the effectiveness of our proposed algorithm in simultaneously reducing users' energy costs is shown via simulations based on realistic system data of California, US.
\end{abstract}	

\newtheorem{definition}{\underline{Definition}}[section]
\newtheorem{fact}{Fact}
\newtheorem{assumption}{Assumption}
\newtheorem{theorem}{\underline{Theorem}}[section]
\newtheorem{lemma}{\underline{Lemma}}[section]
\newtheorem{corollary}{\underline{Corollary}}[section]
\newtheorem{proposition}{\underline{Proposition}}[section]
\newtheorem{example}{\underline{Example}}[section]
\newtheorem{remark}{\underline{Remark}}[section]
\newtheorem{algorithm}{\underline{Algorithm}}[section]
\newcommand{\mv}[1]{\mbox{\boldmath{$ #1 $}}}
\section{Introduction}
The still growing global electricity demand, greenhouse gas emissions from fossil fuel power plants, and climate change have given rise to an increasing combination of renewable energy sources such as solar and wind into the grid. By integrating renewable energy generators at the user level, the demand of individual users can be met locally, which effectively reduces both the carbon dioxide emissions of fossil fuel based power plants and the transmission losses from power plants to the end-users. 
Moreover,  demand response (DR) capability of users helps  smooth out the inherently variable power output from fluctuating renewable energy resources and improve the system reliability \cite{Palomar,Aghaei,Vedady}. 
Specifically, DR can adjust users' power consumption over time to match the power generation of their individual renewable energy generators as closely as possible. This reduces the need of purchasing power from the grid, especially during the peak-demand period where the electricity price is high, which in fact reduces the need for peak power plants on the generation side. 

In practice, DR may not be sufficient to alleviate the intermittent  and stochastic nature of renewable energy generators, since users have also ``must-run'' loads such as lighting that cannot be deferred. In this case, energy storage systems (ESSs) can be deployed to help users by being charged whenever there is renewable energy surplus and discharged in case of energy deficit. Thanks to  the technology advances, integrating ESSs at the user level, e.g., residential and commercial users, has become viable \cite{Tesla}. 
However, due to high upfront investment cost (especially for the large number of users and without  sufficient  government funding) as well as the space limitation,  installing distributed ESSs for individual users  may not be feasible in all circumstances. Consequently, the concept of {\it shared ESS} has become appealing \cite{Paridari,Z.Wang}, by which the surplus renewable energy of some users can be charged into a shared  ESS, and then be discharged by others upon their renewable energy deficit. In general, the shared ESS can be considered as a third party, where users can sell/buy energy to/from it when needed, but at lower costs compared to the tariffs offered by the grid. 

In this paper,  a system with multiple {\it self-interested} users, each with renewable energy generation, fixed  and controllable loads, and one  ESS shared among all users is considered. It is  assumed that renewable energy generation at all users can be perfectly predicted. 
In practice,  each  user needs  to  be motivated  to sell/buy  energy to/from the shared ESS.  
To do so,  an iterative algorithm is proposed  by which a central controller coordinates the energy charged/discharged to/from the shared ESS such that users'  individual energy costs reduce simultaneously as compared to the case where they do not trade energy with the shared ESS. 
The proposed algorithm works based on only limited information received from each user and thus preserves their privacy. 
Next, given the optimized charging/discharging values, each user independently optimizes the energy consumption of its controllable loads and that purchased from the grid. 
For performance benchmark, the  case of {\it cooperative} users that all belong to the same entity or different entities with common interests is also studied. In this case, all users seek a common goal, e.g., minimization of the total energy cost, and  thus  share all the required information with the central controller. 
Finally, the performance of the proposed algorithms for the cases of self-interested and cooperative users  are evaluated  using simulations based on realistic system data of California, US \cite{Load_Profile,NREL,NREL_Solar,Price_CAISO}.
 
The energy management problem for users with ESSs has been well studied in the literature.  However, most of the previous works, e.g., \cite{Palomar,Zhang,Koutsopoulos,Katayoun_journal}, assume that either each user owns an ESS that is not shared with others or all users have common interests and cooperate to follow a common goal, e.g., minimizing the total energy cost, where their individual energy costs and privacy issues are not considered. On the other hand, the idea of a  shared ESS among users and network operator was introduced in \cite{Z.Wang}, and interesting preliminary results were reported. The proposed policy for charging/discharging and satisfying the demand responsive loads in \cite{Z.Wang} makes decisions heuristically and  based on only the hourly electricity prices offered by the grid operator, while other practical considerations are neglected.  Moreover, users are considered to be cooperative; thus, the cost reduction of individual user in using the shared ESS is not discussed.  
Recently, \cite{Paridari} solved the cost minimization problem for energy consumers with demand response capability with no renewable energy integration. Their proposed distributed algorithm aims at minimizing the total energy cost of all users and the resulting benefit in cost reduction is then fairly shared among users according to their flexibility in  load shifting.

In this paper, we design an algorithm by leveraging convex optimization techniques to solve the energy management problem of self-interested users sharing a common ESS while exchanging only limited information with the central controller. In contrast to the prior works on the management of users with a shared ESS \cite{Z.Wang,Paridari}, our proposed algorithm always guarantees that energy costs of individual users reduce concurrently compared to the case of no energy trading with the shared ESS, which motivates users to participate in the energy  trading  program in practice.\vspace{-0.1cm}
\section{System Model}
\begin{figure}[t!]
\centering
\includegraphics[width=6.8cm]{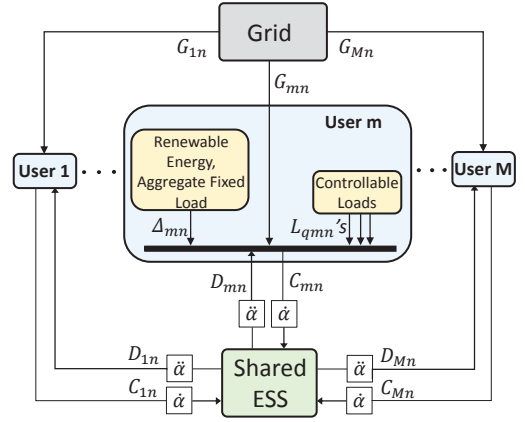}\\ \vspace{-.1cm}
\caption{Illustration of our considered system with multiple self-interested energy users and a shared   ESS.}\label{fig:SystemModel}\vspace{-.3cm}
\end{figure}

\vspace{-.1cm}
As shown in Fig. \ref{fig:SystemModel}, we consider a system of  $M > 1$ users, indexed by $m$, $m \in {\cal M} \in \{1,\ldots,M\}$, each of which can be a single energy consumer (residential, commercial, or industrial) or a group of consumers controlled by an aggregator.
Specifically, we consider that users have their own  renewable energy generators supplying a part or all of their loads over time. The users' loads are divided into two main types; fixed and controllable,  where fixed loads need to be satisfied at the instant requested, while controllable loads can be satisfied within   given desired time periods.  
An energy storage system (ESS) is shared among all users, where they  charge/discharge to/from it whenever necessary. The users  receive/pay money when charging/discharging to/from the shared ESS, with a priori known prices.
We consider a central controller  coordinates the use of the shared ESS based on the information received from  users, and optimizes the energy charged/discharged to/from the shared ESS by all users such that their energy costs decrease at the same time   compared to the case that they operate independently without charging/discharging to/from the shared ESS. 
Furthermore, we consider that users are all connected to the grid, which consists of conventional fossil fuel based power plants, and
can draw energy from it in case of renewable energy deficit.  Last, for our convenience, we use a time-slotted system with slot index $n$, $n \in {\cal N}=\{1,\ldots,N\}$, with $N > 1$ denoting the total number of scheduling times slots, where the duration of each slot is normalized to a unit time and power and energy are thus used interchangeably in this paper.
In the following, we define each system component in  detail.
\subsubsection{Grid Energy Cost} Let $G_{mn}\ge 0$ denote the energy drawn from the grid by user $m$ at time slot $n$, where its corresponding cost for the user is modelled by $f_{mn}(G_{mn}) \ge 0$. We assume that $f_{mn}(G_{mn})$  is a convex and monotonically increasing function over $G_{mn}\ge 0$. Note that our proposed grid energy cost model is time-varying in general.

\subsubsection{Shared ESS} Let $0 \le C_{mn} \le \overline{C}$ and $0 \le D_{mn} \le \overline{D}$ denote the energy charged/discharged to/from the shared ESS by  user $m$ at time slot $n$, respectively, where $\overline{C}>0$ and $\overline{D}>0$ are the maximum charging and discharging rates of the shared ESS, respectively. The energy losses during the charging and discharging processes are specified by charging and discharging efficiency parameters, denoted by $0< \dot{\alpha} < 1$ and $0 <\ddot{\alpha} < 1$, respectively. We denote $S_n\ge 0$ as the available energy in the shared ESS at the beginning of time slot $n$,  which can be derived recursively as follows: \vspace{-.2cm}
\begin{align} \label{eq:storage-State}
S_{n+1}=S_n+\dot{\alpha} \sum_{m=1}^{M}{C_{mn}} - \frac{1}{\ddot{\alpha}} \sum_{m=1}^{M}{D_{mn}}.
\end{align}
Furthermore, practical ESS  has a finite capacity and cannot be  fully discharged to avoid deep discharging. We thus  have the following constraints for the states of the shared ESS:\vspace{-.15cm}
\begin{align}\label{eq:storage}
\underline{S} \le S_{n} \le \overline{S},~\forall n \in {\cal N}, 
\end{align}
where $\underline{S} > 0$ and $\overline{S} > \underline{S}$ are the minimum and maximum   allowed states of the shared ESS, respectively. We set $\underline{S} \le S_{1} \le \overline{S}$ by default. Intuitively, it is not optimal for users to charge/discharge to/from the shared ESS at the
same time due to the energy loss in charging/discharging processes; hence, for each user we  consider $\{C_{mn}\}$ and $\{D_{mn}\}$ satisfying \vspace{-.15cm}
\begin{align}
C_{mn}D_{mn}=0, ~\forall n\in {\cal N}. \label{const:zeroCD}
\end{align}
Last, we assume that users  receive/pay money when charging/discharging to/from the shared ESS, with  known prices. Let $\dot{\omega}_{mn}> 0$ and $\ddot{\omega}_{mn}> 0$,  denote the selling/buying prices of user $m$ to/from the shared ESS at time slot $n$, which are set lower than the  price offered by the grid at the same time slot. This  motivates the energy trading with the shared ESS.

\subsubsection{Controllable Loads} Let $Q_m\ge 1$ denote the number of controllable loads of user $m$, indexed by $q$, $q \in {\cal Q}_m=\{1,\ldots,Q_m\}$. Specifically, controllable load $q$ of user $m$ requires $E_{qm}>0$ amount of energy to complete its task  over the given  time slots $\underline{n}_{qm}\le n \le \overline{n}_{qm}$, with $1\le \underline{n}_{qm}<N$  and  $\underline{n}_{qm}< \overline{n}_{qm}\le N$ denoting  the  starting and termination time slots, respectively. For convenience, we define ${\cal N}_ {qm}=\{\underline{n}_{qm},\ldots,\overline{n}_{qm} \}$. Let $L_{qmn}\ge 0$ denote  the energy allocated to controllable load $q$ of user $m$ at time slot $n$.
Due to practical considerations, over time slots $n \in {\cal N}_{qm}$, $L_{qmn}$ should be higher (lower) than a given minimum (maximum) threshold
$\underline{L}_{qm}> 0$ ($\overline{L}_{qm}> \underline{L}_{qm}$). However, over time slots $n \not \in {\cal N}_{qm}$, $L_{qmn}=0$, since the load should be satisfied  just within its given  scheduling time period ${\cal N}_{qm}$.  By default, we set $(\overline{n}_{qm}-\underline{n}_{qm})\underline{L}_{qmn} <E_{qm} < (\overline{n}_{qm}-\underline{n}_{qm})\overline{L}_{qmn}$, to ensure that controllable load $q$ of user $m$ is  practically schedulable. To summarize, we have the following constraints for all controllable loads of user $m$:\vspace{-.15cm}\footnote{For simplicity, we assume that the energy consumption of demand responsive loads can be {\it continuously} changed over time.}
\begin{align}
&\sum_{n=\underline{n}_{qm}}^{\overline{n}_{qm}}{{L}_{qmn}}=E_{qm},~\forall q \in {\cal Q}_m, \label{const_Load_1}\\
&\underline{L}_{qm}\le {L}_{qmn} \le \overline{L}_{qm},~\forall n \in {\cal N}_{qm},~\forall q \in {\cal Q}_m \label{const_Load_2}\\
&{L}_{qmn}=0,~\forall n \not \in {\cal N}_{qm},~ \forall q\in {\cal Q}_m\label{const_Load_3}
\end{align}
\subsubsection{Net Energy Profile} Let $R_{mn}\ge 0$ and $\hat{L}_{mn}\ge 0$ denote the renewable energy generation and the aggregate fixed loads  of user $m$ at time slot $n$, respectively. We then define $\Delta_{mn}=R_{mn}-\hat{L}_{mn}$, $n\in \cal N$, as the {\it net energy profile} of user $m$ over time. Generally, $\Delta_{mn}$'s are all stochastic due to the randomness in the renewable energy generation, but can be predicted with finite errors. However, in this paper, we assume that $\Delta_{mn}$'s are perfectly known to individual users prior to the scheduling, e.g., day-ahead energy management. 

Last, since the user $m$ needs to satisfy its combined load, i.e., sum of fixed and controllable load, over each time slot $n$, we consider the following energy neutralization constraints:\vspace{-.15cm}
\begin{align}
G_{mn}-C_{mn}+D_{mn}+\Delta_{mn}\hspace{-.5mm}-\hspace{-.5mm}\sum_{q=1}^{Q_m}{L_{qmn}} \ge 0, ~\forall n\in {\cal N}. \label{const_neut} 
\end{align}

In the following, we formulate the energy management problem for self-interested users. 
 
\section{Problem Formulation}
Let $\mv{c}=[\mv{c}_1^T ~\ldots~ \mv{c}_M^T]^T$ and $\mv{d}=[\mv{d}_1^T~ \ldots~ \mv{d}_M^T]^T$, with $\mv{c}_m=[C_{m1} ~\ldots~ C_{mN}]^T$ and $\mv{d}_m=[D_{m1} ~\ldots~ D_{mN}]^T$, denote the charging and discharging vectors.  Given any charging and discharging vectors $\mv{c}$ and $\mv{d}$ satisfying the practical constraints of the shared ESS given in  (\ref{eq:storage}) and (\ref{const:zeroCD}), the minimum energy cost of user $m$, denoted by $F_m(\mv{c},\mv{d})$, is derived as\vspace{-.15cm}
\begin{align*}
&\mathrm{(P1-m)}: \\ 
 &F_m(\mv{c},\mv{d})= \mathop{\mathtt{min}}_{{\cal  X}_m} {\sum_{n=1}^{N}{ f_{mn}(G_{mn})-\dot{\omega}_{mn} C_{mn}+\ddot{\omega}_{mn} D_{mn} }} \\ 
&~~~~~~~~~~~~~~~\mathtt{s.t.} 
~(\ref{const_Load_1})-(\ref{const_neut}),\\
&~~~~~~~~~~~~~~~~~~~~~G_{mn}\ge 0,~ \forall n \in {\cal N},
\end{align*}
where ${\cal X}_m \triangleq \{G_{mn},~L_{qmn},\forall q \in {\cal Q}_m,~\forall n \in {\cal N}\}$ denotes the set of all decision variables for (P1$-$m). It can be readily verified that $F_m(\mv{c},\mv{d})$ is jointly convex  over $\mv{c}$ and $\mv{d}$ \cite{Boyd}. Define  $\Delta\mv{c}=[\Delta\mv{c}_1^T ~\ldots~ \Delta\mv{c}_M^T]^T$ and $\Delta\mv{d}=[\Delta\mv{d}_1^T ~\ldots~ \Delta\mv{d}_M^T]^T$, where $\Delta\mv{c}_m=[\Delta C_{m1} ~\ldots~ \Delta C_{mN}]^T$ and $\Delta\mv{d}_m=[\Delta D_{m1} ~\ldots~ \Delta D_{mN}]^T$. Given any charging and discharging vectors $\mv{c}$ and $\mv{d}$, the energy costs of all users can be decreased simultaneously, compared to the case of no shared ESS, if and only if there exists sufficiently small $\Delta\mv{c}$ and $\Delta\mv{d}$ with $\mv{c}+\Delta\mv{c}$ and $\mv{d}+\Delta\mv{d}$ satisfying $0 \le \mv{c}+\Delta\mv{c} \le \overline{C}$, $0 \le \mv{d}+\Delta\mv{d} \le \overline{D}$, (\ref{eq:storage}), and (\ref{const:zeroCD}) such that $F_m(\mv{c}+\Delta\mv{c},\mv{d}+\Delta\mv{d}) < F_m(\mv{c},\mv{d})$, $\forall m$. In the following, we first characterize the effect of changing the charging and discharging vectors $\mv{c}$ and $\mv{d}$  to  $\mv{c}+\Delta\mv{c}$ and $\mv{d}+\Delta\mv{d}$ on the cost of each individual user $m$, i.e., 
$F_m(\mv{c}+\Delta\mv{c},\mv{d}+\Delta\mv{d})-F_m(\mv{c},\mv{d})$, by investigating the dual problem of (P1$-$m).  Next, we search for desirable $\Delta\mv{c}$ and $\Delta\mv{d}$  that can decrease the energy costs of all users simultaneously.  

Let $\lambda_{n} \ge 0$ be the Lagrange dual variables corresponding to constraints in (\ref{const_neut}).  
The Lagrangian of (P1$-$m)  is given by \vspace{-.15cm}
\begin{align}
&\hspace{-2mm}\mathcal{L}=\sum_{n=1}^{N}{\bigg(f_{mn}(G_{mn})-\lambda_nG_{mn}+\lambda_n\sum_{q=1}^{Q_m}{L_{qmn}}\bigg)}\nonumber\\ &\hspace{-3mm}+\sum_{n=1}^{N}{\hspace{-1mm}\bigg(\hspace{-1mm}(\lambda_n-\dot{\omega}_{mn}) C_{mn}-(\lambda_n-\ddot{\omega}_{mn}) D_{mn} - \lambda_n \Delta_{mn}\hspace{-1mm}\bigg)}.
\end{align} 
The dual function of (P1$-$m) is then given by\vspace{-.15cm}
\begin{align}
{\cal G}=\mathop{\mathtt{min}}_{{\cal X}_m} &~\mathcal{L}\nonumber \\ 
\mathtt{s.t.} 
&~ (\ref{const_Load_1})-(\ref{const_neut}),\nonumber\\
&~G_{mn}\ge 0,~ \forall n \in {\cal N}.
\end{align}
The dual problem of (P1$-$m) is thus expressed as\vspace{-.10cm}
\begin{align}
\mathrm{(D1)}: ~\mathop{\mathtt{max}}_{\{\lambda_n \ge 0\}_{n \in \cal N}} {\cal G}
\end{align}
Denote $\{\lambda_n^{\star} \ge 0,~\forall n\in \cal N\} \subseteq \cal U$ as the optimal solution to (D1), where ${\cal U}$ is the set of all optimal dual variables. Accordingly, $F_m(\mv{c}+\Delta\mv{c},\mv{d}+\Delta\mv{d})-F_m(\mv{c},\mv{d})$ is given by the following lemma.
\begin{proposition}\label{Derivatives}
Under any given $\Delta\mv{c}$ and $\Delta\mv{d}$, the change in the energy cost of user $m$ by adjusting the energy charged/discharged to/from the shared ESS  is given by\vspace{-.1cm}
\begin{align}\label{eq:first_order_App}
F_m(\mv{c}+\Delta\mv{c},\mv{d}&+\Delta\mv{d})-F_m(\mv{c},\mv{d})=\nonumber\\
&\sum_{n=1}^{N}{(\frac{\partial F_m(\cdot)}{\partial C_{mn}}\Delta C_{mn}+\frac{\partial F_m(\cdot)}{\partial D_{mn}} \Delta D_{mn})} 
\end{align}
where $\Delta C_{mn}$ and $\Delta D_{mn}$ are sufficiently small, $\frac{\partial F_m(\cdot)}{\partial C_{mn}}$ and $\frac{\partial F_m(\cdot)}{\partial D_{mn}}$ are derivatives of $F_m(\cdot)$ with respect to $C_{mn}$ and $D_{mn}$, respectively. Note that $F_m(\cdot)$ may not be differentiable in $C_{mn}$ and $D_{mn}$.  However, it follows from \cite{Horst} that the left and right-partial derivatives of  $F_m(\cdot)$ with respect to $C_{mn}$ and $D_{mn}$ still exist, which can be given as follows:\vspace{-.10cm}
\begin{align}
& \frac{\partial F_m(\cdot)}{\partial C_{mn}^+}\hspace{-.5mm}=\hspace{-.5mm}-\dot{\omega}_m+\max_{\lambda_n^{\star}\in\cal U} {\{\lambda_n^{\star}\}}, ~\frac{\partial F_m(\cdot)}{\partial C_{mn}^-}\hspace{-.5mm}=\hspace{-.5mm}-\dot{\omega}_m+\min_{\lambda_n^{\star}\in\cal U} \{{\lambda_n^{\star}}\},\label{eq_Derivative_C}\\
& \frac{\partial F_m(\cdot)}{\partial D_{mn}^+}\hspace{-.5mm}=\hspace{-.5mm}\ddot{\omega}_m-\min_{\lambda_n^{\star}\in\cal U} \{{\lambda_n^{\star}}\},~\hspace{3mm}\frac{\partial F_m(\cdot)}{\partial D_{mn}^-}\hspace{-.5mm}=\hspace{-.5mm}\ddot{\omega}_m-\max_{\lambda_n^{\star}\in\cal U} \{{\lambda_n^{\star}}\}.\label{eq_Derivative_D}
\end{align} 
\end{proposition}\vspace{-.20cm}

Given the partial derivatives in (\ref{eq_Derivative_C}) and (\ref{eq_Derivative_D}),  $F_m(\cdot)$ can be approximated as follows\footnote{In the case that $F_m(\cdot)$ is differentiable, $\lambda_n$ is unique; thus,  right-partial and  left-partial derivatives are equal.}:\vspace{-0.2cm}
\begin{align}
&F_m(\mv{c},\mv{d})=\sum_{n=1}^{N} {\frac{\partial F_m(\cdot)}{\partial C_{mn}^+}[\Delta C_{m,n}]^+ +\frac{\partial F_m(\cdot)}{\partial D_{mn}^+}[\Delta D_{m,n}]^+}\nonumber\\
&-\sum_{n=1}^{N} {\frac{\partial F_m(\cdot)}{\partial C_{mn}^-}[-\Delta C_{m,n}]^+-\frac{\partial F_m(\cdot)}{\partial D_{mn}^-}[-\Delta D_{m,n}]^+}. 
\end{align}
where $[x]^+\triangleq \max(0,x)$.

Given  Proposition \ref{Derivatives}, we seek for sufficiently small  $\Delta\mv{c}$ and  $\Delta\mv{D}$, with $0 \le \mv{c}+\Delta\mv{c} \le \overline{C}$ and $0 \le\mv{d}+\Delta\mv{d} \le \overline{D}$ satisfying the constraints in (\ref{eq:storage}) and (\ref{const:zeroCD}), such that $F_m(\mv{c}+\Delta\mv{c},\mv{d}+\Delta\mv{d}) < F_m(\mv{c},\mv{d})$, $\forall m$. We investigate the existence of such $\Delta\mv{c}$ and $\Delta\mv{d} $  by solving the following feasibility  problem.\footnote{Users are all coupled to each other due to the shared ESS constraints in (\ref{eq:storage}); hence, $\{\Delta{C}_{mn}\}$ and  $\{\Delta{D}_{mn}\}$ are jointly optimized in (F1).}
\begin{align}
&\mathrm{(F1)}:~\texttt{find}~\{\Delta C_{mn},~ \Delta D_{mn},~\forall m \in {\cal M}, ~\forall n \in {\cal N}\}  \nonumber \\ 
\hspace{-3mm}\mathtt{s.t.}
& |\Delta C_{mn}| \leq \rho,~ |\Delta D_{mn}| \leq \rho,~\forall m, ~\forall n \label{const_feasibility_small}\\
&0\hspace{-.8mm}\le\hspace{-.8mm} C_{mn}\hspace{-.8mm}+\hspace{-.7mm}\Delta C_{mn}\hspace{-.8mm} \le\hspace{-.8mm} \overline{C},\hspace{-.5mm}~ 0\hspace{-.8mm}\le\hspace{-.8mm} D_{mn}\hspace{-.7mm}+\hspace{-.7mm}\Delta D_{mn}\hspace{-.8mm} \le\hspace{-.8mm} \overline{D}\label{const_feasibility_1},\hspace{-.8mm} ~\forall m, n\hspace{-.5mm}\\
& S_n\hspace{-1mm}+\hspace{-.7mm}\hspace{-1.2mm} \sum_{m=1}^{M}\hspace{-.9mm}{\hspace{-.6mm}\dot{\alpha}(\hspace{-.6mm}C_{mn}\hspace{-1.1mm}+\hspace{-.8mm}\Delta C_{mn}\hspace{-.6mm}) \hspace{-.8mm}-\hspace{-1mm} \frac{1}{\ddot{\alpha}} (\hspace{-0.6mm}D_{mn}\hspace{-1.1mm}+\hspace{-1mm}\Delta D_{mn}\hspace{-.5mm})} \hspace{-.7mm} \ge \hspace{-.7mm}\underline{S}, \forall m, n \hspace{-1mm} \label{const_feasibility_3}\\
& S_n\hspace{-1mm}+\hspace{-.7mm}\hspace{-1.2mm} \sum_{m=1}^{M}\hspace{-.9mm}{\hspace{-.6mm}\dot{\alpha}(\hspace{-.6mm}C_{mn}\hspace{-1.1mm}+\hspace{-.8mm}\Delta C_{mn}\hspace{-.6mm}) \hspace{-.8mm}-\hspace{-1mm} \frac{1}{\ddot{\alpha}} (\hspace{-0.6mm}D_{mn}\hspace{-1.1mm}+\hspace{-1mm}\Delta D_{mn}\hspace{-.5mm})} \hspace{-.7mm} \le \hspace{-.7mm}\overline{S}, \forall m, n  \label{const_feasibility_4}\\
& (C_{mn}\hspace{-.6mm}+\hspace{-.6mm}\Delta C_{mn})(D_{mn}\hspace{-.6mm}+\hspace{-.6mm}\Delta D_{mn})\hspace{-.6mm}=\hspace{-.6mm}0, ~\forall m, n  \label{const_feasibility_2}\\
&\sum_{n=1}^{N}{(\frac{\partial F_m}{\partial C_{mn}}\Delta C_{mn}\hspace{-.5mm}+\hspace{-.5mm}\frac{\partial F_m}{\partial D_{mn}} \Delta D_{mn}) } \hspace{-.5mm}< 0, \hspace{-.5mm}~\forall m \hspace{-.5mm} \label{const_feasibility_5}  
\end{align} 
where $\rho > 0$ is a small step size. Particularly, the constraints in (\ref{const_feasibility_small}) restricts each $\Delta\mv{c}$ and $\Delta\mv{d}$ to take small steps, since (\ref{eq:first_order_App}) is only valid in proximity of    $\mv{c}$ and $\mv{d}$. The constraints  in (\ref{const_feasibility_1})-(\ref{const_feasibility_2}) present   practical considerations of the shared ESS. Last, the constraints in (\ref{const_feasibility_5}) ensure that  energy costs of all users decrease  simultaneously when $\mv{c}$ and $\mv{d}$ are changed to $\mv{c}+\Delta\mv{c}$ and $\mv{d}+\Delta\mv{d}$. Note that (F1) is a non-convex optimization problem due to constraints (\ref{const_feasibility_2}) and (\ref{const_feasibility_5}). However, constraints (\ref{const_feasibility_small})-(\ref{const_feasibility_4}) specify a convex set over $\{\Delta C_{mn}\}$ and $\{\Delta D_{mn}\}$. To solve (F1), we  search over the set specified by constraints (\ref{const_feasibility_small})-(\ref{const_feasibility_4}) to find $\{\Delta C_{mn}\}$ and $\{\Delta D_{mn}\}$ that satisfy (\ref{const_feasibility_2}) and (\ref{const_feasibility_5}). 
Our algorithm to design the charging and discharging vectors $\mv{c}$ and $\mv{d}$ is summarized in Table I, as Algorithm $1$.\footnote{Note that  (F1) is not convex; hence, its complexity in general grows exponentially with the number of decision variables. However, a rather small number of users (or aggregators) may share a common ESS  in practice. In this case, the complexity of Algorithm $1$ is of no concern.}
\begin{table}[t!]
	\begin{center} 
			\caption{Algorithm for  the Energy Management of Self-Interested Users}
\vspace{0.01cm}\hrule\vspace{0.06cm} \textbf{Algorithm 1}  \vspace{0.02cm}\hrule 
\begin{itemize} 
\item[a)] Initialize $\mv{c}\gets\mv{0},\mv{d}\gets\mv{0}$,  $\rho>0$, and Flag$\gets0$.
\item[b)] {{\bf While } $\text{Flag} \neq 1$  {\bf do}:}
\begin{itemize}
\item[1)] Given the charging and discharging vectors  $\mv{c}$ and $\mv{d}$, each user $m$ computes $\frac{\partial F_m(\cdot)}{\partial C_{mn}}$ and $\frac{\partial F_m(\cdot)}{\partial D_{mn}}$ using (\ref{eq_Derivative_C}) and (\ref{eq_Derivative_D}), respectively. Derivatives are then sent to the central controller.
\item[1)] Given the received derivatives from all $m$ users, the central controller then searches for $\{\Delta{C}_{mn}\}$ and  $\{\Delta{D}_{mn}\}$ via solving the feasibility problem in (F1).  If (F1) is infeasible, Flag$=1$ is set. Otherwise, the charging and discharging vectors $\mv{c}$ and $\mv{d}$ are updated as $\mv{c}+\Delta\mv{c}$ and $\mv{d}+\Delta\mv{d}$, respectively. 
\end{itemize}
\item[c)] The central controller announces $\{\Delta{C}_{mn}\}$ and  $\{\Delta{D}_{mn}\}$ to each user $m$ as the final decision for charging and discharging. Given the optimized $\{\Delta{C}_{mn}\}$ and  $\{\Delta{D}_{mn}\}$, each user then independently solves (P1-m) to obtain $\{G_{mn}\}$ and $\{L_{qmn}\}$. 
\end{itemize}
\vspace{0cm} \hrule \vspace{0.2cm}
\label{Table:Algorithm}
\end{center}\vspace{-.8cm}
\end{table}

\section{Benchmark: Cooperative Users}
In this section, we consider that all users either belong to the same entity or different entities with common interests and are willing to  share all the required information (including their renewable energy generation, information of controllable loads,  etc.) with the central controller. Given the provided information, the central controller  minimizes the {\it total} energy cost of all users by jointly optimizing the energy charged/discharged to/from the shared ESS, that drawn from the grid, and that consumed by controllable loads.  We thus have the following optimization problem for {\it cooperative} users: \vspace{-.5cm}
\begin{align*}
\mathrm{(P2)}:&  \mathop{\mathtt{min}}_{\cal Y}
~ {\sum_{m=1}^{M}{\sum_{n=1}^{N}{ f_{mn}(G_{mn})-\dot{\omega}_{mn} C_{mn}+\ddot{\omega}_{mn} D_{mn} }}} \\ 
\mathtt{s.t.} 
~&(\ref{eq:storage}),~\textnormal{and}~  (\ref{const_Load_1})-(\ref{const_neut}),~\forall m \in {\cal M}, \\
~& G_{mn}\hspace{-.5mm}\ge\hspace{-.5mm} 0,  0 \hspace{-.5mm} \le\hspace{-.5mm}  C_{mn} \hspace{-.5mm} \le\hspace{-.6mm}  \overline{C}, 0 \hspace{-.5mm} \le \hspace{-.8mm} D_{mn} \hspace{-.5mm} \le\hspace{-.6mm}  \overline{D}, \forall m \hspace{-.6mm} \in\hspace{-.6mm}  {\cal M}, \forall n \hspace{-.6mm} \in\hspace{-.6mm}  {\cal N}\hspace{-.4mm} ,
\end{align*} 
where ${\cal Y}\triangleq\{C_{mn},~D_{mn},~G_{mn},~L_{qmn},~\forall q \in {\cal Q}_m,~\forall m \in {\cal M},~\forall n \in {\cal N}\}$ denotes the set of all decision variables for (P2). 
It can be easily verified that (P2) is a convex optimization problem and can be solved using standard convex optimization techniques such as the interior point method \cite{Boyd}. Herein, we do not provide the optimal closed-form solution to (P2) for brevity and leave it for the journal version of this work. Note that constraints in (\ref{const:zeroCD}) are not explicitly  included in (P2); however,  it can be shown that the optimal solution to (P2) always satisfies these constraints.  

\begin{remark}
In the  energy management for cooperate users, users need  to  share all the required information with the central controller (e.g., the net energy profiles, requirements of controllable loads, etc., over the scheduling  period). However, Algorithm 1 proposed for  self-interested users  can be implemented by only knowing  the partial derivatives of users' energy costs $F_m(\cdot)$'s with respect to $C_{mn}$ and $D_{mn}$, $\forall m\in {\cal M}$, $\forall n \in {\cal N}$ (i.e., $4MN$ scalars in each iteration); hence, the privacy of users is preserved. In addition, Algorithm 1 minimizes energy costs of users simultaneously such that all achieve  lower energy costs compared to the case of no charging/discharging to/from the shared ESS, which is a motivation for self-interested users to trade  energy with the shared ESS.
\end{remark}
\section{Simulation Results}
\begin{figure}[t!]
	\centering
	\includegraphics[width=6.2cm]{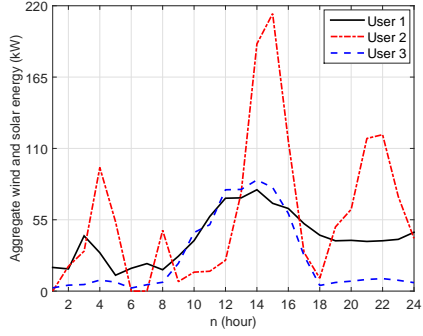}\\ \vspace{-.25cm}
	\caption{Aggregate solar and wind energy generation over one day.}\label{fig:Users_RE}\vspace{-.4cm}
\end{figure}
In this section, we consider a system of three users $M=3$ based on the real data available from California, US, over one day, 5 January, 2006 \cite{Load_Profile,NREL,NREL_Solar,Price_CAISO}. 
We model user 1 as a building with residential consumers, user 2 as a medium-size office, and user 3 as a restaurant \cite{Load_Profile}. We assume that each user has its own renewable energy generators, including both solar and wind, with  profiles shown in Fig. \ref{fig:Users_RE} \cite{NREL,NREL_Solar}. 

\begin{figure}[t!]
	\centering
	\includegraphics[width=6.2cm]{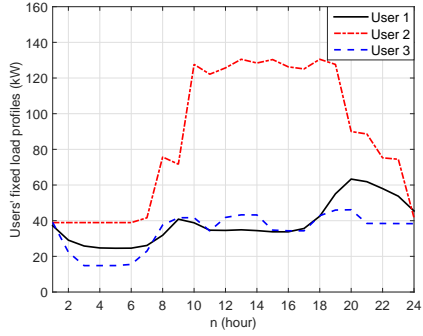}\\ \vspace{-.25cm}
	\caption{Load profiles of the three users over one day.}\label{fig:Users_Load_Profile}\vspace{-.1cm}
\end{figure} 
\begin{table}[t!]
	\centering
	\caption{Users' controllable loads parameters}
	\label{Table:Flexible_Load}
	\begin{tabular}{|c|c|c|c|c|c|}
		\hline
		\multirow{2}{*}{\begin{tabular}[c]{@{}c@{}}Controllable\\ Loads\end{tabular}}       & \multirow{2}{*}{$\underline{n}_{qm}$\hspace{-.5mm}} & \multirow{2}{*}{$\overline{n}_{qm}$\hspace{-.5mm}} & \multirow{2}{*}{$\underline{L}_{qm}$\hspace{-.6mm} (kW)\hspace{-1mm}} & \multirow{2}{*}{$\overline{L}_{qm}$\hspace{-.6mm} (kW)\hspace{-1mm}} & \multirow{2}{*}{$E_{qm}$\hspace{-1mm} (kWh)\hspace{-1mm}} \\
		&                                       &                                      &                                            &                                           &                                 \\ \hline
		\multirow{2}{*}{\begin{tabular}[c]{@{}c@{}}User 1:\\ Electric vehicle\end{tabular}} & \multirow{2}{*}{$1$}                  & \multirow{2}{*}{$9$}                 & \multirow{2}{*}{$0$}                       & \multirow{2}{*}{$20$}                     & \multirow{2}{*}{$50$}           \\
		&                                       &                                      &                                            &                                           &                                 \\ \hline
		\multirow{2}{*}{\begin{tabular}[c]{@{}c@{}}User 2: \\ Air conditioner\end{tabular}} & \multirow{2}{*}{$7$}                  & \multirow{2}{*}{$19$}                & \multirow{2}{*}{$35$}                      & \multirow{2}{*}{$70$}                     & \multirow{2}{*}{$600$}          \\
		&                                       &                                      &                                            &                                           &                                 \\ \hline
		\multirow{2}{*}{\begin{tabular}[c]{@{}c@{}}User 3: \\ Dishwasher\end{tabular}}      & \multirow{2}{*}{$1$}                  & \multirow{2}{*}{$8$}                 & \multirow{2}{*}{$0$}                       & \multirow{2}{*}{$20$}                     & \multirow{2}{*}{$63$}           \\
		&                                       &                                      &                                            &                                           &                                 \\ \hline
	\end{tabular}\vspace{-.3cm}
\end{table}
Profiles of the users' fixed loads are shown in Fig. \ref{fig:Users_Load_Profile}. For simplicity, we assume that each user has only one controllable load. Specifically,  user 1 has an electric vehicle (EV), user 2 is equipped with smart air conditioner system that can adjust its temperature set point  to manage its energy consumption over time, and user 3 has a smart commercial  dishwasher that its energy consumption can be controlled. Details of the users' controllable loads are given in Table \ref{Table:Flexible_Load}. It is shown that the EV needs to be charged from 00:00 AM to 8:00 AM (time slots $1\le n \le 9$) and receive the total energy of $50$ kWh during this period. The energy consumption of the air conditioner of user 2 can be modified during the office hour from 7:00 AM to 8:00 PM (time slots $7 \le n \le 19$) while satisfying a minimum power consumption of $35$ kW; thus, it cannot be completely turned off to avoid user's inconvenience. Finally, the dishwasher of user 3 can operate flexibly during 00:00 AM to 7:00 AM (time slots $1\le n \le 8$) and consume $63$ kWh  during this time \cite{Dishwasher}.
\begin{figure}[t!]
	\centering
\includegraphics[width=9cm]{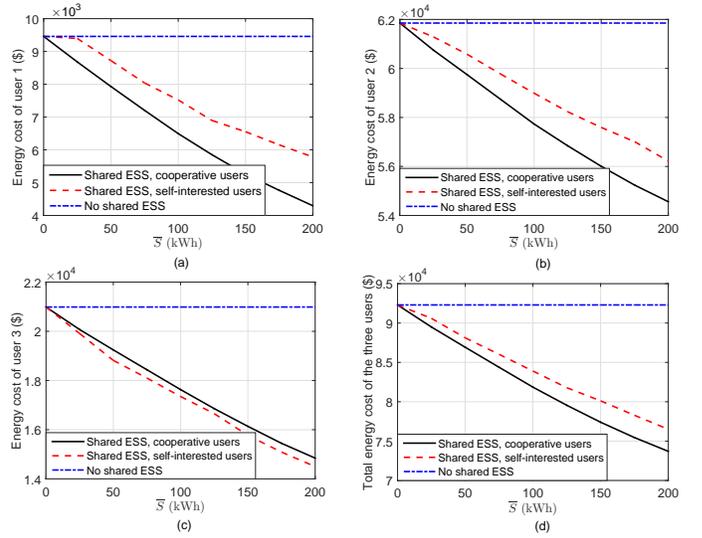}\\ \vspace{-.2cm}
	\caption{Users' energy costs over the capacity of the shared ESS: a) User 1, b) User 2, c) User 3,  d) Total energy cost of the three users.}\label{fig:Cost_Comparison}\vspace{-.3cm}
\end{figure}
\begin{figure}[t!]
	\centering
	\includegraphics[width=6.2cm]{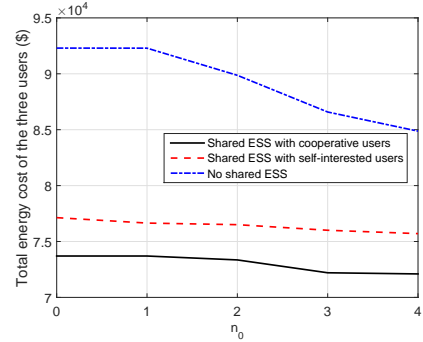}\\ \vspace{-.25cm}
	\caption{Total energy cost of all three users over termination time of controllable loads.}\label{fig:Load_Flexib}\vspace{-.2cm}
\end{figure}

For the shared ESS, we consider  sodium-sulfur based batteries with maximum and minimum capacities of $\overline{S}=200$ kWh and  $\underline{S}=0.1\overline{S}$, respectively,  charging and discharging efficiencies of $\dot{\alpha}=\ddot{\alpha}=0.87$, and maximum charging and discharging rates of $\overline{C}=\overline{D}=0.15\overline{S}$ \cite{SS_Battery}. We also set $S_1=\underline{S}$.  
Last, we  model the cost function of purchasing energy from the  grid as $f_{mn}=45G_{mn}$ \cite{Wood}.

The energy cost of each user and the total energy of all three users under three different cases of no shared ESS, cooperative, and self-interested users are shown in Figs. \ref{fig:Cost_Comparison} (a)-(d). It is observed from Fig. \ref{fig:Cost_Comparison} (d) that the case of no shared ESS results in the highest total energy cost, while the case of cooperative users has  the lowest total energy energy cost, since all users share the required information with the central controller and the charging/discharging, purchasing from the  grid, and the energy consumption of controllable loads are jointly optimized to minimize the total energy cost. The results in Fig. \ref{fig:Cost_Comparison} (d) also show that with only limited information sharing in the case of self-interested users, the total energy cost reduces remarkably as compared to the case of no shared ESS, while performs close to the lower bound derived from the energy management for cooperative users. Note  that the energy cost of user 3 in the case of cooperative users is higher than that  of the case of self-interested users, as shown in  \ref{fig:Cost_Comparison} (c). This can be justified due to fact that the  total energy cost resulting from the energy management of cooperative users  always achieves the  lowest energy cost, while for individual users, this may not always be  the case.


Next, by setting $\overline{S}=200$ kWh, we investigate the impact of the flexibility of controllable loads on the total energy cost. Specifically, we assume that the termination time of users' controllable loads can be extended by $0\le n_0 \le 4$ time slots, i.e., we set  $\overline{n}_{qm}=\overline{n}_{qm}+n_0$, $\forall q \in {\cal Q}_m$, $\forall m \in {\cal M}$.   
Fig. \ref{fig:Load_Flexib} plots the resulted total energy cost over  $n_0$, from which it is observed that extending the termination time of controllable loads by only four time slots, i.e., $n_0=4$, can highly reduce the total energy cost in the case of no shared ESS, while for the other two cases of shared ESS with self-interested and cooperative users the cost reduction is negligible. This is due to the fact that in the absence of the shared ESS, energy deficit is  satisfied by optimizing the consumption of  controllable loads  and/or purchasing energy from the grid. As a result, the higher flexibility of controllable loads in this case can substantially reduce the need for purchasing  energy from the grid. However, in the presence of the shared ESS, users can also trade with it to deal with energy deficit and thus the flexibility of the controllable loads becomes less effective. 
\begin{table}[]
	\centering
	\caption{Increment of energy cost in absence of renewable energy generators of one user}\vspace{-.1cm}
	\label{Table:RE_Omit}
	\begin{tabular}{|c|c|c|c|}
		\hline
		\multirow{2}{*}{\begin{tabular}[c]{@{}c@{}}\hspace{-1mm}Renewable energy \\ Integration\end{tabular}\hspace{-1mm}} & \multicolumn{3}{c|}{Energy cost increase ($\%$)}                      \\ \cline{2-4} 
		& \hspace{-.55mm}Cooperative  \hspace{-.55mm} & \hspace{-.55mm}Self-interested  \hspace{-.55mm} & \hspace{-.55mm}Without shared ESS\hspace{-.55mm} \\ \hline
		Users 1 and 2                                                                        & $34.01$           & $30.97$               & $20.15$            \\ \hline
		Users 1 and   3                                                                        & $82.17$           & $77.45$               & $60.47$            \\ \hline
		Users 2 and   3                                                                        & $56.74$           & $52.93$               & $37.30$            \\ \hline
	\end{tabular}\vspace{-.2cm}
\end{table}
%

Table \ref{Table:RE_Omit} shows the increment in the total energy cost resulting from eliminating the renewable energy generators in one user, either 1, 2, or 3, compared to the case that all have renewable energy generation.  It is  observed that eliminating renewable energy sources of one user in the case that they do not share an ESS results in a lower energy cost increase percentage   compared to the other two cases of cooperative and self-interested users with a shared ESS. This is because when users do not share an ESS, they operate independently (cannot have energy sharing via the shared ESS) and zero energy generation of each user will affect only that specific user. In contrast, the cases of cooperative and self-interested users  have energy cooperation via the shared ESS and the lack of renewable energy generation of one user affects the total system energy cost more substantially. 
\section{Conclusion}
In this paper, we have studied the energy management problem for self-interested users with renewable energy integration, DR capability, and a shared ESS. We proposed an iterative algorithm that gradually updates the energy charged/discharged to/from the shared ESS by all users to reduce their individual energy costs simultaneously  compared to the case of no energy trading with the shared ESS while sharing only limited information with the central controller. Our simulation results show that by deploying our proposed energy management algorithm, all users can achieve much lower energy costs  compared to the case of no energy trading with the shared ESS  and the total energy cost in this case performs fairly close to the lower bound derived from the energy management of fully cooperative users. Our model is useful in practical systems where installing individual ESS for each user is either very costly or requires space that is not available. Note that this paper does not provide  algorithms for the real-time energy management of the self-interested users in the presence of stochastic renewable energy generation. Devising efficient online algorithms is a possible future direction of this work.

\end{document}